\documentstyle[twoside,fleqn,espcrc2]{article}
\input{epsf.sty}

\def\slash{{/\!\!\!}}

\newcommand{\AmS}{{\protect\the\textfont2
  A\kern-.1667em\lower.5ex\hbox{M}\kern-.125emS}}

\title{On tadpole improvement for staggered fermions}

\author{Maarten Golterman\address{Department of Physics, 
        Washington University, St. Louis, MO 63130, USA}}%
       
\begin{document}

\begin{abstract}
An explanation is proposed for the fact that 
Lepage--Mackenzie tadpole improvement does not
work well for staggered fermions.  The idea appears to work for all
renormalization constants which appear in the staggered fermion self-energy.
Wilson fermions are also discussed.
\end{abstract}

\maketitle

 
One-loop renormalization constants for a number of staggered-fermion
operators have large finite parts
(see {\it e.g.} ref.\ \cite{Patel}), which appear not to be explained
by gluon-tadpole contributions.  This is in contrast to the situation
with Wilson fermions, where tadpole improvement \cite{Lepage} appears
to work well.

Staggered and Wilson fermions differ in the way they deal with the
doubling problem: in the Wilson case, they get a mass of the order of
the cutoff and decouple,
while they are present in the staggered case, yielding a number of
continuum flavors which is a multiple of four.
The staggered doublers do contribute to loop diagrams, and here I suggest
that this may explain why tadpole improvement does not work well for
staggered fermions, by considering the one-loop fermion self-energy.

Let us first consider the Wilson case.  The one-loop self-energy is
\cite{Karsten}
\begin{eqnarray}
\Sigma(p)&=&\frac{4}{3}g^2\Bigl[-\frac{1}{8\pi^2}
\int_0^1 dx\;(i\slash p(1-x)+2m) \nonumber \\
&&\times\log{(x(1-x)p^2+xm^2)} \label{sew} \\
&&+\sigma_0+i\slash p\sigma_1+m\sigma_2\Bigr]\;, \nonumber
\end{eqnarray}
with $\sigma_{0,1,2}$ given in Table~\ref{tab:wsigmas} (in Feynman
gauge; 1st column).  The 2nd column gives the tadpole-improved values;
the 3rd the contribution to those of the 2nd from the integration region
$-\pi/2<\ell_\mu\le\pi/2$ ($\ell$ is the loop momentum).

\begin{table*}[htb]
\setlength{\tabcolsep}{1.5pc}
\newlength{\digitwidth} \settowidth{\digitwidth}{\rm 0}
\catcode`?=\active \def?{\kern\digitwidth}
\begin{tabular*}{\textwidth}{@{}l@{\extracolsep{\fill}}rrr}
\hline
& \multicolumn{1}{r}{no improvement}
& \multicolumn{1}{r}{tadpole improved}
& \multicolumn{1}{r}{reduced BZ}  \\
\hline
$\sigma_0$ & $0.326?$ & $0.0158$ & $0.0169??$ \\
$\sigma_1$ & $-0.0878$ & $-0.0103$ & $-0.00839^*$ \\
$\sigma_2$ & $-0.0120$ & $-0.0120$ & $-0.0184??$ \\
\hline
\end{tabular*}
\caption{Wilson contact terms.
\null$^*$ Depends on routing of external momentum, which
can make a $\sim 25$\% difference.}
\label{tab:wsigmas}
\end{table*}

The table shows that gluon-tadpole improvement works well (I used
the mean link in Feynman gauge).  If we split up the Brillouin zone
(BZ) for $\ell$ as $\ell=\pi_A+\tilde\ell$, $A=1,\dots,16$, with
$\tilde\ell\in (-\pi/2,\pi/2]$ and $\pi_A\in\{(0,0,0,0),(\pi,0,0,0),
\dots\}$, we see that most of the tadpole-improved values comes from
the region with $\pi_A=0$.  The doublers do not contribute much, since
they are suppressed by the Wilson mass term.

Now, let us consider staggered fermions.  The most general four-flavor
mass matrix is
\begin{eqnarray}
M&=&m^S+m^V_\mu\xi_\mu+\frac{1}{2}m^T_{\mu\nu}(-i\xi_\mu\xi_\nu)
\nonumber \\
&&+m^A_{5\mu}i\xi_\mu\xi_5+m^P_5\xi_5\;, \label{mass}
\end{eqnarray}
where the $\xi_\mu$ are a set of gamma matrices in flavor space
($m^T_{\mu\nu}$ is antisymmetric).  The labels $S$, $V$, {\it etc.}
(for scalar, vector, {\it etc.}) denote irreducible
representations of the staggered
fermion symmetry group, and correspond to zero-, one-, {\it etc.}
link operators in the lattice action.

The staggered one-loop
self-energy is \cite{Golterman}
\begin{eqnarray}
\Sigma(p)&=&\frac{4}{3}g^2\Bigl[-\frac{1}{8\pi^2}U^\dagger\!
\int_0^1\!\! dx\;(i\slash p(1-x)+2M_d) \nonumber \\
&&\times\log{(x(1-x)p^2+xM_d^2)}\;U \nonumber \\
&&+\tau\;i\slash p+\sigma_S\;m^S+\sigma_V\;m^V_\mu\xi_\mu \label{ses} \\
&&+\frac{1}{2}\sigma_T\;m^T_{\mu\nu}(-i\xi_\mu\xi_\nu)
+\sigma_A\;m^A_{5\mu}i\xi_\mu\xi_5 \nonumber \\
&&+\sigma_P\;m^P_5\xi_5\Bigr]\;, \nonumber
\end{eqnarray}
with $U$ diagonalizing $M$: $M_d=UMU^\dagger$.
The values of $\tau$ and the $\sigma$'s
are given in Table~\ref{tab:ssigmas} (${\overline\sigma}_V$
is a similar constant for a one-link mass term for ``reduced"
staggered fermions \cite{Golterman}).

\begin{table*}
\setlength{\tabcolsep}{1.5pc}
\settowidth{\digitwidth}{\rm 0}
\catcode`?=\active \def?{\kern\digitwidth}
\begin{tabular*}{\textwidth}{@{}l@{\extracolsep{\fill}}rrr}
\hline
& \multicolumn{1}{r}{no improvement} 
& \multicolumn{1}{r}{tadpole improved}
& \multicolumn{1}{r}{reduced BZ} \\
\hline
$\tau$ & $-0.0446?$ & $ 0.0329$ & $0.0149^*?$ \\
$\sigma_S$ & $0.197??$ & $0.197?$ & $0.0276??$ \\
$\sigma_V$ & $0.00385$ & $0.0813$ & $0.0244??$ \\
$\sigma_T$ & $-0.117??$ & $0.0380$ & $0.0111??$ \\
$\sigma_A$ & $-0.209??$ & $0.0233$ & $0.00516?$ \\
$\sigma_P$ & $-0.294??$ & $0.0161$ & $0.000294$ \\
${\overline\sigma}_V$ & $0.0813?$ & $0.159?$ & $0.0332??$ \\
\hline
\end{tabular*}
\caption{Staggered contact terms. \null$^*$ See Table~\ref{tab:wsigmas}.}
\label{tab:ssigmas}
\end{table*}

For the staggered case we see that gluon-tadpole improvement does not work
for $\sigma_S$, $\sigma_V$ and ${\overline\sigma}_V$ (it does work for
$\sigma_T$, $\sigma_A$ and $\sigma_P$), and that the
$\pi_A=0$ region does not give the main contribution, giving a value
typically much smaller than the tadpole-improved value.

The interpretation I would like to put forward is as follows.  Consider
the standard one-loop self-energy diagram:
\begin{figure}[h]
\includegraphics{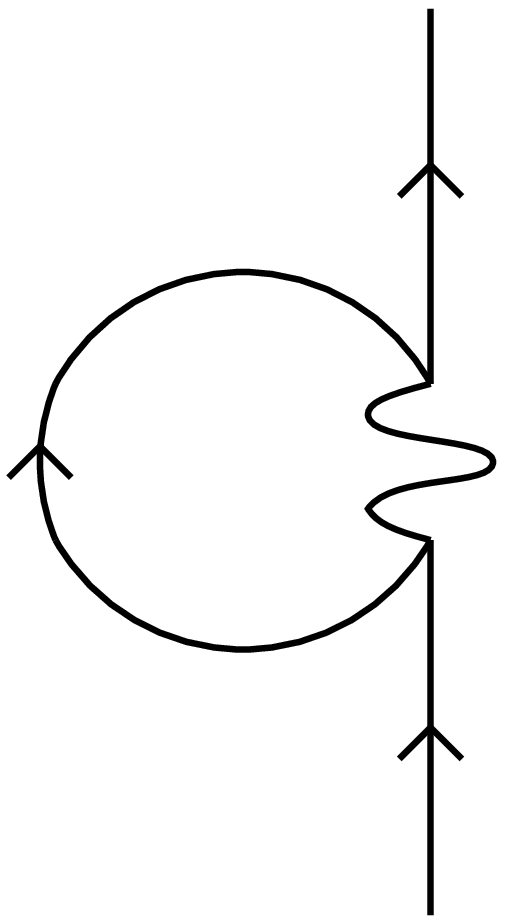}
\vspace*{1.3cm}
\end{figure}

\noindent
For $\pi_A\ne 0$, the gluon propagator is very suppressed (of order
$a^2$), and the gluon line in the diagram is effectively reduced to
a four-fermion coupling between the staggered flavors.  The fermion
propagator, however, has poles for $\pi_A\ne 0$, and the diagram
contributes ``doubler-tadpoles" to the finite part of the one-loop
self-energy.  If we contract the gluon line to a point, the integrand
near the doubler poles ($\pi_A\ne 0$) goes like $1/\tilde\ell^2$, and
produces contributions very much like the usual gluon tadpoles: the
integral is quadratically divergent ($1/a^2$), with an extra factor
$\sim a^2$ coming from the contracted gluon propagator.

This idea can be tested by replacing the gluon propagator 
$\Bigl[(4\sum_\mu\sin^2{(\ell_\mu/2)}\Bigr]^{-1}$ (in Feynman gauge) by
\begin{eqnarray}
&&\frac{1}{4}[P^-(\ell_1) P^+(\ell_2) P^+(\ell_3) P^+(\ell_4)+\dots] 
\label{ff} \\
&&\frac{1}{8}[P^-(\ell_1) P^-(\ell_2) P^+(\ell_3) P^+(\ell_4)
+\dots] \nonumber \\
&&\frac{1}{12}[P^-(\ell_1) P^-(\ell_2) P^-(\ell_3) P^+(\ell_4)
+\dots] \nonumber \\
&&\frac{1}{16}P^-(\ell_1) P^-(\ell_2) P^-(\ell_3) P^-(\ell_4)
\;, \nonumber
\end{eqnarray}
where
\begin{equation}
P^\pm(\ell_\mu)\equiv\frac{1}{2}(1\pm\cos{\ell_\mu})\;. \label{proj}
\end{equation}
The factors $P^-(\ell_1)P^+(\ell_2)P^+(\ell_3)P^+(\ell_4)$,
{\it etc.} are ``smooth projectors" onto the regions $\pi_A+
\tilde\ell$ with $\pi_A\ne 0$ of the Brillouin zone.  The smoothness
makes it possible to interpret this ``gluon" exchange as a local
four-fermion operator.  The fractions $1/4$, $1/8$, {\it etc.}
are the values of the real gluon propagator at $\ell=\pi_A\ne 0$.
Note that the region around $\pi_A=0$ is ``projected" onto 0.

With this replacement, {\it i.e.} with this four-fermion interaction,
one obtains the results of Table~\ref{tab:ffsigmas}.
\begin{table*}[htb]
\setlength{\tabcolsep}{1.5pc}
\settowidth{\digitwidth}{\rm 0}
\catcode`?=\active \def?{\kern\digitwidth}
\begin{tabular*}{\textwidth}{@{}l@{\extracolsep{\fill}}rrr}
\hline
& \multicolumn{1}{r}{tadpole improved}
& \multicolumn{1}{r}{four-fermion}
& \multicolumn{1}{r}{fraction} \\
\hline
$\tau$ & $0.0329$ & $0.0185$ & $0.56$ \\
$\sigma_S$ & $0.197?$ & $0.179?$ & $0.91$ \\
$\sigma_V$ & $0.0813$ & $0.0668$ & $0.82$ \\
$\sigma_T$ & $0.0380$ & $0.0309$ & $0.81$ \\
$\sigma_A$ & $0.0233$ & $0.0237$ & $1.0?$ \\
$\sigma_P$ & $0.0161$ & $0.0227$ & $1.4?$ \\
${\overline\sigma}_V$ & $0.159?$ & $0.134?$ & $0.84$ \\
\hline
\end{tabular*}
\caption{Staggered contact terms -- comparison with four-fermion values.}
\label{tab:ffsigmas}
\end{table*}
We note that
\begin{itemize}
\item the four-fermion constants reproduce the tadpole-improved
contact terms quite well, especially the larger ones;
\item gluon-tadpole improvement is also needed;
\item for Wilson fermions, the ``four-fermion values" for $\sigma_0$,
$\sigma_1$ and $\sigma_2$ are $-0.00222$, $-0.00111$ and $0.00949$,
respectively, to be compared with the gluon-tadpole improved values
given in Table~\ref{tab:wsigmas}. 
\end{itemize}

The idea presented here can be checked on the many other staggered-fermion
renormalization constants that have been calculated to one loop in
perturbation theory.  Also, since the approach is gauge-dependent, it
should be checked for other gauges, such as Landau gauge.

Assuming that the idea is correct, one can ask how better estimates
of staggered-fermion renormalization constants can be obtained.
This could be done by taking the staggered-fermion
theory with only the four-fermion interactions (no gluons), and
computing the renormalization constants in this theory numerically.
(These constants are finite, because the four-fermion interactions
are irrelevant operators, proportional to $g^2$.)

To one-loop, one then multiplies these by 
the perturbatively calculated constants of
the full theory (with gluons), with the four-fermion part taken out,
and the appropriate power of the mean link for gluon-tadpole improvement.
For example, for the wave-function renormalization, we would get
\begin{equation}
Z_2=u_0^{-1}Z_2^{\rm 4f}\left[1-\frac{4}{3}g^2\!
\left(\!-\frac{1}{8\pi^2}\log{a\mu}+\Delta\tau\right)\right],
\label{wfr}
\end{equation}
where $u_0$ is the mean link, $Z_2^{\rm 4f}$ is the wave-function
renormalization of the four-fermion theory, and $\Delta\tau=
0.0144$ is the difference between the first two numbers of 
Table~\ref{tab:ffsigmas}.

This procedure resembles gluon-tadpole improvement, in that it
partially resums the perturbative expansion, and it is equally
heuristic.  A disadvantage is that a numerical computation is 
needed in the four-fermion theory for each operator.  
A complete nonperturbative determination in the full theory
\cite{Ishizuka} may therefore be preferable not only in principle,
but also in practice.

A different approach would be to consider improved actions for 
lattice QCD with staggered fermions (see {\it e.g.} ref.\ \cite{Toussaint}).
For improved actions, one expects the couplings of high-momentum
gluons and fermions to be smaller than in the unimproved case, 
which would presumably lead to
smaller finite parts of the renormalization constants.

\medskip
I would like to thank Jim Hetrick, Mike Ogilvie, Steve Sharpe and
Doug Toussaint for discussions.  This work is supported in part by
the US Dept. of Energy under the Outstanding Junior Investigator
program.

\end{document}